\documentclass[english,english,aps,prd,amsmath,amssymb,superscriptaddress]{revtex4}
\usepackage[T1]{fontenc}
\usepackage[latin9]{inputenc}
\usepackage{amsbsy}
\usepackage{esint}

\makeatletter
\@ifundefined{textcolor}{}
{%
 \definecolor{BLACK}{gray}{0}
 \definecolor{WHITE}{gray}{1}
 \definecolor{RED}{rgb}{1,0,0}
 \definecolor{GREEN}{rgb}{0,1,0}
 \definecolor{BLUE}{rgb}{0,0,1}
 \definecolor{CYAN}{cmyk}{1,0,0,0}
 \definecolor{MAGENTA}{cmyk}{0,1,0,0}
 \definecolor{YELLOW}{cmyk}{0,0,1,0}
 }

\makeatother

\usepackage{babel}

\begin{document}

\title{Magnetic Susceptibility in Strongly Coupled Systems}

\author{M.J.Luo}

\affiliation{Department of Modern Physics, University of Science and Technology
of China, Anhui 230026, People's Republic of China}
\begin{abstract}
We study the magnetic susceptibility at large 't Hooft coupling by
computing the correlation function of the magnetizations in the strongly
coupled Maxwell theory in large-N limit with finite temperature and
chemical potential, within the framework of the AdS/CFT correspondence.
We show that in strong coupling limit the magnetic susceptibility
is independent to the temperature and be universal, measured in the
unit of magnetic permeability of the bulk space. A comparison with
the weak coupling system, the Pauli paramagnetic susceptibility, is
also discussed. 
\end{abstract}
\maketitle

\section{Introduction}

Magnetic susceptibility $\chi$ is an important property of materials
indicating the magnetic response to an applied external magnetic field.
In microscopic picture, $\chi$ measures the propagation of the
collective wave mode of local magnetization $\mathbf{M}$, called
spin wave. Its quanta are the magnons, up to a gyromagnetic ratio constant,
the magnetic susceptibility and the spin-spin correlation function can be viewed as one thing. In fact, the magnetic susceptibility
is the magnetization or spin transport. Like other theoretical studies
of transport coefficients, e.g. see \cite{Jeon:1995zm,Arnold:2000dr},
the magnetic susceptibility could be calculated by using the standard
perturbative technique in quantum field theory, which is based on
the precondition that the magnetizations or, equivalently, the spins
taken by constituent fermions interact weakly with each other. The
condition is fulfilled in the Fermi-liquid theory, which is a theory
in the vicinity of a trivial fixed point \cite{Anderson:1990}. When
the interaction becomes stronger, the calculations are notoriously
difficult. However, it is conjectured that another non-trivial fixed
point exists \cite{PhysRevD.58.046004} and corresponds to a strongly
coupled conformal fields theory that is dual to a string/M theory
in an AdS space, the AdS/CFT correspondence \cite{Maldacena:1997re,Gubser:1998bc,Witten:1998qj}.
These two fundamentally different fixed points correspond to the extremely
weak and strong coupling limit \cite{Anderson:1990}.

An example of the study of the magnetic susceptibility in extremely
weak coupling limit is the Pauli paramagnetism, which describes $\chi$ in electron gas \cite{pauli_paramag}. The validity of the
weakly interacting description is based on the fact that the Coulomb
interactions are effectively screened, so the Coulomb interaction
becomes a short range force characterized by the Debye mass. But it is known that the magnetic
interactions which are mediated by the magnons can not be effectively
screened and spoil the normal Fermi-liquid behavior of the system
\cite{PhysRevLett.74.1423}, therefore, the non-perturbative effects
in magnetic susceptibility are thought to be important, especially in
strongly coupled system. In the extremely strong coupling limit, the
't Hooft coupling tends to infinity, the string/M theory is reduced
to a classical (super)gravity, so it allows us to do calculation of
the correlation functions in the limit.

One of the famous predictions \cite{Policastro:2001yc,Policastro:2002se}
of the AdS/CFT correspondence
was that the ratio of the shear viscosity $\eta$ to the entropy density
$s$, in $\mathcal{N}=4$ super-Yang-Mills (SYM) theory at large 't
Hooft coupling with finite temperature, equals to $1/4\pi$ (in natural
units), it is a universal quantity independent with the microscopic
details, which agrees well with the observation from the strongly coupled
quark-gluon-plasma produced in relativistic heavy ion collision \cite{PhysRevLett.97.162302}. The strongly coupled QCD theory is
an important area for applying the results. The surprising success of the AdS/CFT correspondence at low energies is probably the result of the universality in its predictions. Note that the magnetic susceptibility
is dimensionless, so a naive guess is that it may be universal as
well in the prediction from the approach, it is very interesting to
check this idea by detail calculations. Another motivation for doing
the calculation is that the topic of magnetic aspects of the quark
matter in the phase diagram of QCD have attracted many interests,
e.g. see \cite{Tatsumi:2008nx,Tatsumi:2008gu,Agasian:2008tb,PhysRevD.78.074033,Kharzeev:2009pj,Kharzeev:2010gd}.

In this paper, we will work in the framework of AdS/CFT correspondence
at large 't Hooft coupling limit with finite temperature and chemical
potential \cite{Ge:2008ak}. The idea to calculate the magnetic susceptibility
in this framework is simple. Similarly to the standard procedure in calculating
the other transport coefficients in Minkowski prescription \cite{Son:2002sd},
one places the magnetization $M_{i}$ on the 4-dimensional boundary
that couples to the magnetic field $H_{i}$ which propagates in the
5-dimension bulk AdS space. One can write down the action of the
magnetic fields in the bulk space deduced from the Maxwell action
in the AdS background. Depending on the thermodynamical variables
of the system, we need to place a Schwarzschild or charged Reissner-Nordstrom
black hole into the AdS space, which corresponds to introduce finite
temperature and/or chemical potential, respectively. The two point
correlation function (in Minkowski space) of $M_{i}$ can be computed
by performing the functional derivative with respect to the magnetic
field $H_{i}$ as a source on the boundary. Here the magnetic fields
$H_{i}$ in the bulk are dual to $M_{i}$, which is analogous to the
case that we compute the correlation function of charged currents $J_{i}$
where the gauge fields $A_{i}$ in the bulk are dual to $J_{i}$.

The paper is organized as follows. In section \ref{sec:Preliminaries}
we briefly define the magnetic susceptibility from the linear response
theory and review the computing framework of the Green's functions
from AdS/CFT correspondence in Minkowski prescription. In section
\ref{sec:Holographic-calculation}, we perform a detail calculation
to the magnetic susceptibility in two cases, the system with temperature
and temperature together with chemical potential. We also compare
our result with the one computed from the weakly coupled limit, the
Pauli paramagnetic susceptibility. Section \ref{sec:Conclusion} contains
the conclusions.

\section{\label{sec:Preliminaries}Preliminaries}

\subsection{Magnetic susceptibility in the linear response theory}

In this section, we set up a field theoretical framework for the response
of a system at equilibrium to small perturbations. The framework allows
us to relate a two point correlation function of magnetizations to
the magnetic susceptibility of the system.

Let us consider the response of the system to the presence of a weak external
magnetic field $H^{i}(x)$ which couples to a the magnetization $M_{i}$.
Then the Hamiltonian is perturbed by a term

\begin{equation}
\delta\mathcal{H}=\int d^{4}xM_{i}(x)H^{i}(x),\end{equation}
where the index $i=1,2,3$. The standard perturbation theory in textbook
of quantum mechanics tells us that it produces a change in the expectation
value of the operators\begin{equation}
\delta\langle M_{i}(x)\rangle=\int d^{4}x^{\prime}\tilde{G}_{ij}^{R}(x,x^{\prime})H^{j}(x^{\prime})+\mathcal{O}(H^{2}),\end{equation}
in which\begin{equation}
\tilde{G}_{ij}^{R}(x,x^{\prime})=-i\theta(t-t^{\prime})\langle[M_{i}(x),M_{j}(x^{\prime})]\rangle\end{equation}
is the retarded Green's function. The result can also be found by
using the Kubo formula, which tells us that to first order in the
time-dependent perturbation, the induced vector current (here it is
the perturbative wave of magnetization $M_{i}(x)$, or the spin wave
current) is equal to retarded correlator to the vector current with
the perturbation evaluated in equilibrium. The Fourier transformed
linear response then takes a simple form\begin{equation}
\delta M_{i}(k)=G_{ij}^{R}(k,0)H_{j}(k)+\mathcal{O}(H^{2}),\label{eq:linear_perturbation}\end{equation}
where the Fourier transformation of the retarded Green's function
is\begin{equation}
G_{ij}^{R}(k)=\int d^{4}xe^{-ikx}\tilde{G}_{ij}^{R}(x,0).\end{equation}

To see the relation between the retarded Green's function to the magnetic
susceptibility $\chi_{ij}$, we write down its definition \begin{equation}
M_{i}(k)\equiv\chi_{ij}(k)H_{j}(k),\end{equation}
which $\chi_{ij}$ is a second rank tensor. Compared with Eq.(\ref{eq:linear_perturbation}),
consider that here the external perturbation is weak, at linear level,
the magnetic susceptibility tensor is identified with the retarded
Green's function, i.e. the two point magnetization-magnetization correlation
function \begin{equation}
\chi_{ij}(k)=G_{ij}^{R}(k,0).\label{eq:chi_green}\end{equation}

\subsection{Minkowski correlators in AdS/CFT correspondence}

In order to calculate the two point magnetization-magnetization correlation
function of a thermal strongly coupled system in Minkowski space,
one need to discuss in detail a prescription for computing a two-point
Green's function from gravity, followed by the AdS/CFT correspondence.
One can write the AdS/CFT correspondence as the equality in Euclidean
version\begin{equation}
\langle e^{\int_{\partial\mathcal{M}}M_{i}H_{0}^{i}}\rangle=e^{-S_{cl}[H]}.\end{equation}
The left hand side is a generating functional for the correlators
of magnetization in the boundary field theory, which is conjectured
as a $\mathcal{N}=4$ SU(N) SYM theory at large N limit. When the
't Hooft coupling $g_{YM}^{2}N$ tends to infinity, the right hand
side tends to the action of the classical Einstein (super)gravity,
and the external magnetic field $H$ propagates in the bulk $AdS_{d+1}$
space, with its boundary condition $H_{0}$ couples to the magnetization
$M_{i}$ on the boundary $\partial\mathcal{M}$ of the AdS space.
In order to introduce a finite temperature to the system, one has
to place a black hole to the AdS space, the metric in Minkowski version
can be written as

\begin{equation}
ds^{2}=\frac{(\pi TR)^{2}}{u}\left(-f(u)dt^{2}+dx^{2}+dy^{2}+dz^{2}\right)+\frac{R^{2}}{4u^{2}f(u)}du^{2},\label{eq:background}\end{equation}
where for Schwarzschild-AdS background we have $f(z)=1-u^{2}$ and
$u=r_{0}^{2}/r^{2}$, $r_{0}$ is the radius of the horizon of the
black hole, in which $T=r_{0}/\pi R^{2}$ is the Hawking temperature,
the horizon locates at $u=1$, the boundary at $u=0$. 

As proposed by Son and Starinets \cite{Son:2002sd}, to generalize
the AdS/CFT correspondence from the Euclidean to Minkowski version,
formally we have the relation\begin{equation}
\langle e^{i\int_{\partial\mathcal{M}}M_{i}H_{0}^{i}}\rangle=e^{iS_{cl}[H]},\label{eq:correspondence}\end{equation}
together with the incoming-wave boundary condition at the horizon,
i.e. all modes are absorbed into the black hole horizon but no ones
can emit. By using the Eq.(\ref{eq:chi_green}) and Eq.(\ref{eq:correspondence}),
the retarded Green's function, and then the magnetic susceptibility
in a strongly coupled system can be computed from the second functional
derivative of $S_{cl}$ with respect to the boundary value $H_{0}$,\begin{equation}
\chi_{ij}=-2\frac{\delta^{2}S_{cl}[H]}{\delta H_{0}^{i}\delta H_{0}^{j}}\Biggl|_{u\rightarrow0}.\label{eq:chi_from_action}\end{equation}

\section{\label{sec:Holographic-calculation}Holographic calculation}

\subsection{Finite Temperature}

In this section, we work on the 5-dimensional Schwarzschild-AdS background
and consider the perturbations of magnetic field $H^{i}$ in it. Our
starting point is the 5-dimensional Maxwell action in the background
Eq.(\ref{eq:background}),

\begin{equation}
S=-\frac{1}{4g_{YM}^{2}}\int d^{5}x\sqrt{-g}F_{\mu\nu}F^{\mu\nu},\end{equation}
where \begin{equation}
g_{YM}^{2}=16\pi^{2}R/N^{2}\label{eq:gYM}\end{equation}
is the coupling constant. In this paper, what we are interested in
is the correlator of magnetizations coupled to the magnetic fields
which are directly observed physical quantities unlike the gauge
potential $A_{\mu}$, so we will use the electric and magnetic fields
($E_{i}$,$H_{i}$) as fundamental dynamical variables. One can rewrite
the action as\begin{equation}
S=-\frac{1}{2g_{YM}^{2}}\int d^{5}x\sqrt{-g}\left(\epsilon_{0}E_{i}E^{i}-\mu_{0}H_{i}H^{i}\right),\end{equation}
where $\epsilon_{0}$ is the electric permittivity, $\mu_{0}$ the
magnetic permeability of the vacuum in the bulk space. Here we assume
that the backreaction of the source on the boundary to the bulk electromagnetic
fields is small, the electric and magnetic wave that will propagate
in the bulk along $u$ axis are almost purely transverse, so we shall
set $E_{u}=H_{u}=0$, the physical independent fields are those with
index $i=x,y,z=1,2,3$. The physical components are defined as \[
\sqrt{\epsilon_{0}}E^{i}=F^{i0},\quad\sqrt{\mu_{0}}H^{i}=\frac{1}{2}\epsilon^{ijk}F_{jk},\]
where $\epsilon^{ijk}=1$ for that the order of indices $(ijk)$ are
an even/odd permutation of $(123)$. One can use the Fourier decomposition\begin{eqnarray}
P_{i}(x,u) & = & \int\frac{d^{4}K}{(2\pi)^{4}}e^{-i\omega t+ik\cdot x}P_{i}(K,u),\quad P=H\;\mathrm{or}\; E.\end{eqnarray}
By locally using the Maxwell equation in 4-dimensional space\begin{eqnarray}
\nabla\times\mathbf{E} & = & -\mu_{0}\frac{\partial\mathbf{H}}{\partial t},\end{eqnarray}
to replace the transverse electric fields $E_{i}$ with magnetic fields
$H_{i}$ locally. Then the action can be written as\begin{equation}
S=-\frac{1}{2g_{YM}^{2}}\mu_{0}\int du\int\frac{d^{4}K}{(2\pi)^{4}}\sqrt{-g}\left(\epsilon_{0}\mu_{0}\omega^{2}-k^{2}\right)\frac{1}{k^{2}}H_{i}(K,u)H^{i}(K,u).\end{equation}
Without loss of generality, one can set the speed of light $c^{2}=(\epsilon_{0}\mu_{0})^{-1}=1$
in the bulk space, so we have\begin{equation}
S=\frac{1}{2g_{YM}^{2}}\mu_{0}\int du\int\frac{d^{4}K}{(2\pi)^{4}}\sqrt{-g}\frac{K^{2}}{k^{2}}H_{i}(K,u)H^{i}(K,u),\label{eq:action}\end{equation}
where $K^{2}=-\omega^{2}+k^{2}$, we denote $K_{\mu}=(\omega,\mathbf{k})$
locally as a 4-momentum. The magnetic fields can be decomposed as\begin{equation}
H^{i}(K,u)=h_{K}^{i}(u)H_{0}^{i}(K),\label{eq:decomposition}\end{equation}
note that $h_{K}^{i}$ equals to 1 at the boundary $u=0$, \begin{equation}
\lim_{u\rightarrow0}h_{K}^{i}(u)=1.\label{eq:boundary_condition}\end{equation}
The equations of motion of magnetic fields $H_{i}(K,u)$ in the extra
dimension $u$ are given by the decoupled equations of motion of $h_{K}^{i}(u)$\begin{equation}
\frac{1}{\sqrt{-g}}\partial_{u}\left(\sqrt{-g}g^{uu}\partial_{u}h_{K}^{i}\right)-g^{\mu\nu}K_{\mu}K_{\nu}h_{K}^{i}=0.\label{eq:EOM}\end{equation}
Introducing dimensionless energy and momentum in unit of temperature
\begin{equation}
\boldsymbol{\omega}=\frac{\omega}{2\pi T},\quad\boldsymbol{k}_{i}=\frac{k_{i}}{2\pi T},\end{equation}
substituting the metric Eq.(\ref{eq:background}) into Eq.(\ref{eq:EOM}),
we have\begin{equation}
\left(h_{K}^{i}\right)^{\prime\prime}+\left(\frac{f^{\prime}}{f}-\frac{1}{u}\right)\left(h_{K}^{i}\right)^{\prime}+\left(\frac{\boldsymbol{\omega}^{2}}{uf^{2}}-\frac{\boldsymbol{k}^{2}}{uf}\right)h_{K}^{i}=0,\label{eq:diff_equ}\end{equation}
in which the prime stands for the derivative with respect to $u$.
The Eq.(\ref{eq:diff_equ}) is a second-order differential equation
for $h_{K}^{i}(u)$ in which at the horizon $u=1$ is a singular point,
and behaves as $h_{K}^{i}=(1-u)^{\nu}F^{i}(u)$, where $F^{i}(u)$
is a regular function. There are only two values of $\nu_{\pm}=\pm i\boldsymbol{\omega}/2$,
and the incoming wave boundary condition at the horizon is $\nu_{-}$.
Then we obtain the equation for $F^{i}(u)$,\begin{equation}
F^{\prime\prime}+\left(-\frac{1+u^{2}}{uf}+\frac{i\boldsymbol{\omega}}{1-u}\right)F^{\prime}+\left(-\frac{i\boldsymbol{\omega}}{2uf}\right)F+\frac{\boldsymbol{\omega}^{2}\left[4-u(1+u)^{2}\right]}{4uf^{2}}F-\frac{\boldsymbol{k}^{2}}{uf}F=0,\end{equation}
Since the three equations are decoupled and identical, we have omitted
the superscript $i$ and denoted the solution as $F$. In the low
frequency and long wavelength limit, the $\boldsymbol{\omega}$ and
$\boldsymbol{k}$ can be considered small, we solve the equation perturbatively
by expanding the solution $F$ in powers of these small parameters\begin{equation}
F(u)=F_{0}+\boldsymbol{\omega}F_{1}+\boldsymbol{k}^{2}G_{1}+\boldsymbol{\omega}^{2}F_{2}+\boldsymbol{\omega}\boldsymbol{k}^{2}H_{1}+...\end{equation}
The leading order contribution is given by first three functions $F_{0},F_{1},G_{1}$,
which can be solved explicitly, the integration constants can be fixed
by requiring that these functions are regular at the horizon $u=1$,
and vanish in the limit $u\rightarrow1$ (except $F_{0}$). We obtain
\begin{equation}
F_{0}=C,\quad F_{1}=-\frac{iC}{2}\log\frac{1+u}{2},\quad G_{1}=-C\log\frac{1+u}{2}.\end{equation}
The constant $C$ is determined by the boundary condition Eq.(\ref{eq:boundary_condition}),
so we have\begin{equation}
C=\frac{1}{1+(\frac{i}{2}\boldsymbol{\omega}+\boldsymbol{k}^{2})\log2}.\end{equation}
Near the boundary, the radial derivative of the field behaves as\begin{equation}
\lim_{u\rightarrow0}\partial_{u}h_{K}=-\boldsymbol{k}^{2}-\frac{\boldsymbol{\omega}^{2}}{4}\log2+\frac{i}{2}\boldsymbol{\omega}\boldsymbol{k}^{2}\log2.\end{equation}
at leading order\begin{equation}
\lim_{u\rightarrow0}\partial_{u}h_{K}=-\boldsymbol{k}^{2}+...\label{eq:h_prime}\end{equation}
where ... denotes the higher order corrections, $\mathcal{O}(\boldsymbol{\omega}^{2})$
and $\mathcal{O}(\boldsymbol{\omega}\boldsymbol{k}^{2})$. Substituting
the solution into Eq.(\ref{eq:action}) and Eq.(\ref{eq:decomposition})
and integrate $u$ by part, we get\begin{equation}
S=\frac{1}{2g_{YM}^{2}}\mu_{0}\int\frac{d^{4}K}{(2\pi)^{4}}\sqrt{-g}\frac{1}{k^{2}}H_{0}^{i}g^{ij}\left[g^{uu}h_{-K}^{i}\partial_{u}h_{K}^{j}\right]H_{0}^{j}.\end{equation}
So according to the Eq.(\ref{eq:chi_from_action}) and Eq.(\ref{eq:gYM}),
we have\begin{equation}
G_{ij}^{R}=\frac{N^{2}\delta_{ij}}{32\pi^{2}}\mu_{0}+...\end{equation}
We see that the correlation functions is isotropic, the magnetic susceptibility
of the system can be written as a scalar\begin{equation}
\chi=\frac{N^{2}}{32\pi^{2}}\mu_{0}+...\end{equation}

\subsection{Finite Temperature and Chemical potential}

To generalize this result to a system with finite density, one need
to replace the Schwarzschild black hole by a charged black hole, namely,
the Reissner-Nordstrom-AdS (RN-AdS) background, which has the same form
as Eq.(\ref{eq:background}) with a different structure of horizon
\begin{equation}
f(u)=(1-u)(1+u-au^{2}),\label{eq:f_RN}\end{equation}
where $a$ is parameter that relates to the charge of the black hole.
The temperature and chemical potential of the system can be now written
as\begin{equation}
T=\frac{1}{2\pi b}(1-\frac{a}{2}),\quad\Sigma=\frac{1}{2b}\sqrt{\frac{3a}{2}},\end{equation}
in which $b$ is another parameter related to the mass of the black
hole \cite{Ge:2008ak}. The calculating process is similar, we need
to solve the differential equations Eq.(\ref{eq:diff_equ}) by using
Eq.(\ref{eq:f_RN}). Similarly, the solution is found to be\begin{equation}
h_{K}(u)=C(1-u)^{-i\boldsymbol{\omega}/2}\left[1+\boldsymbol{\omega}F_{1}+\boldsymbol{k}^{2}G_{1}+\mathcal{O}(\boldsymbol{\omega}^{2},\boldsymbol{\omega}\boldsymbol{k}^{2}...)\right],\end{equation}
with\begin{eqnarray}
C & = & \frac{1}{1-\frac{1}{4}\boldsymbol{\omega}\left[\pi+i\log(a-2)\right]-\frac{3i\boldsymbol{\omega}+4\boldsymbol{k}^{2}}{2\sqrt{-1-4a}}\left[\tan^{-1}\left(\frac{2a-1}{\sqrt{-1-4a}}\right)+\tan^{-1}\left(\frac{1}{\sqrt{-1-4a}}\right)\right]},\\
F_{1} & = & -\frac{i}{4}\log\left(\frac{2-a}{1+u-au^{2}}\right)+\frac{3i}{2\sqrt{-1-4a}}\left[\tan^{-1}\left(\frac{2au-1}{\sqrt{-1-4a}}\right)-\tan^{-1}\left(\frac{2a-1}{\sqrt{-1-4a}}\right)\right],\\
G_{1} & = & \frac{2}{\sqrt{-1-4a}}\left[\tan^{-1}\left(\frac{2au-1}{\sqrt{-1-4a}}\right)-\tan^{-1}\left(\frac{2a-1}{\sqrt{-1-4a}}\right)\right].\end{eqnarray}
The behavior near the boundary is\begin{equation}
\lim_{u\rightarrow0}\partial_{u}h_{K}=-\boldsymbol{k}^{2}.\end{equation}
Differ from Eq.(\ref{eq:h_prime}), there are no higher order corrections
such as $\mathcal{O}(\boldsymbol{\omega}^{2})$, $\mathcal{O}(\boldsymbol{\omega}\boldsymbol{k}^{2})$.
Applying the prescription formulated in the previous section, one
finds\begin{equation}
\chi=\frac{N^{2}}{32\pi^{2}}\mu_{0},\label{eq:result}\end{equation}
which is our final result for the magnetic susceptibility at large
't Hooft coupling $g_{YM}^{2}N\gg1$. It can be regarded as a nontrivial
prediction from the strongly coupled $\mathcal{N}=4$ SYM theory at
finite temperature and chemical potential. The first observation is
that in this limit $\chi$ is independent with the temperature and
the 't Hooft coupling, it is so simple and be a universal quantity.
It is measured in the unit of the magnetic permeability $\mu_{0}$
of the bulk space. 

The result is positivity, if we have an analytic continuation for
the result from large N to finite N, the system would be paramagnetic.
Note that in the weak coupling limit, the quasi-particle gas is paramagnetic,
the Pauli paramagnetism \cite{pauli_paramag}, it is interesting to
compare the Eq.(\ref{eq:result}) with the Pauli paramagnetic susceptibility.
In this weak coupling regime, the $\chi_{Pauli}$ comes from the contribution
of free quasi-particles near the Fermi surface \cite{pauli_paramag}\begin{equation}
\chi_{Pauli}=\mu_{0}\mu_{B}^{2}\rho,\end{equation}
where $\mu_{0}$ is the vacuum permeability, $\mu_{B}^{2}=g_{YM}^{2}N/4m^{2}$
is the Bohr magneton, $m$ the effective mass of the quasi-particle,
and \begin{equation}
\rho=-2N\int\frac{d^{3}k}{(2\pi)^{3}}\frac{\partial n_{k}}{\partial\omega_{k}}=\frac{Nk_{F}m}{\pi^{2}},\end{equation}
is the density of states near the Fermi surface, where $k_{F}$ is
the fermi momentum, $N$ the number of species of the fermions. Finally
we get\begin{equation}
\chi_{Pauli}=\frac{N}{4\pi^{2}}\mu_{0}\frac{(g_{YM}^{2}N)k_{F}}{m},\quad g_{YM}^{2}N\ll1.\label{eq:chi_pauli}\end{equation}
The Eq.(\ref{eq:result}) and Eq.(\ref{eq:chi_pauli}) implies that
in strong coupling regime the {}``quasi-particle'' (if we can still
denote them by this name) becomes heavy so that the effecitve mass
is comparable to the order of the numerator near the Fermi surface,
i.e. $m\sim\mathcal{O}(g_{YM}^{2}k_{F})$, and $m=8g_{YM}^{2}k_{F}$
for $g_{YM}^{2}N\rightarrow\infty$. Note that the life-time of quasi-particle
is $\tau\sim1/m$, so the notion of the long-lived quasi-particle at
the Fermi surface does not hold any more in the strongly coupled system,
instead of a broadened spectral density and/or smeared Fermi surface,
which has been observed in the studies on strongly coupled non-Fermi-liquid
system \cite{Lee:2008xf,Liu:2009dm,Faulkner:2010da,Faulkner:2011tm}.
We expect that $\chi$ behaves similarly in a non-Fermi-liquid system.

\section{\label{sec:Conclusion}Conclusion}

In this paper, we have calculated the real-time correlation function
of the magnetization $\mathbf{M}$, i.e. the magnetic susceptibility
$\chi$, in the $\mathcal{N}=4$ SYM theory at finite temperature
and chemical potential by using the Minkowski AdS/CFT prescription.
We show that in extremely strong coupling limit, the magnetic susceptibility,
measured in the unit of magnetic permeability in the bulk space,
does not vary with the temperature and 't Hooft coupling. It is found
to be universal and independent from the microscopic details. We expect that our result can be extended and applied to the strongly coupled
quark-gluon-plasma and the non-Fermi-liquid system observed in strange
metal phase of cuperate superconductors and many heavy fermion materials.

\end{document}